\documentclass[a4paper]{jpconf}
\usepackage{graphicx}
\begin{document}

\title{Evidence of the QCD tricritical endpoint existence at NICA-FAIR energies}

\author{K. A. Bugaev$^1$, R. Emaus$^2$, V. V. Sagun$^{1,3}$, 
A. I. Ivanytskyi$^1$,  L. V. Bravina$^2$, D. B. Blaschke$^{4,5,6}$, 
E. G. Nikonov$^7$,  A. V. Taranenko$^6$, E. E. Zabrodin$^{2,6,8}$ and 
G. M. Zinovjev$^1$ }

\address{$^1$ Bogolyubov Institute for Theoretical Physics, Metrologichna str. 14$^B$, Kiev 03680, Ukraine}
\address{$^2$ Department of Physics, University of Oslo, PB 1048 Blindern, 
N-0316 Oslo, Norway}
\address{$^3$ CENTRA, Instituto Superior T$\acute{e}$cnico, Universidade de Lisboa,
Av. Rovisco Pais 1, 1049-001 Lisboa, Portugal}
\address{$^4$ Institute of Theoretical Physics, University of Wroclaw, pl. M. Borna 9, 50-204 Wroclaw, Poland}
\address{$^5$ Bogoliubov Laboratory of Theoretical Physics, JINR Dubna, Joliot-Curie str. 6, 141980 Dubna, Russia}
\address{$^6$ National Research Nuclear University ``MEPhI'' (Moscow Engineering Physics
Institute), Kashirskoe Shosse 31, 115409 Moscow, Russia}
\address{$^7$   Laboratory for Information Technologies, JINR, Joliot-Curie str. 6, 141980 Dubna, Russia}
\address{$^8$ Skobeltzyn Institute of Nuclear Physics, Moscow State University,
119899 Moscow, Russia}

\ead{Bugaev@th.physik.uni-frankfurt.de}

\begin{abstract}
We present a summary of possible signals of the chiral symmetry restoration and deconfinement phase transitions  which may be, respectively, probed at the center of mass collision energies at  4.3-4.9 GeV and above 8.7-9.2 GeV. It is argued that these signals may evidence for an existence of the tricritical endpoint of QCD phase diagram at the collision energy around  8.7-9.2 GeV. The equation  of state  of  hadronic matter with the restored chiral symmetry is discussed and the number of bosonic and fermionic degrees of freedom is found.
\end{abstract}

\section{Introduction}
One of the most important directions of  physics of heavy ion collisions  is related to a location of the 
(tri)critical endpoint ((3)CEP) of the  quantum chromodynamics  (QCD) phase diagram. At present the lattice QCD 
cannot tell us whether at high baryonic charge densities  the chiral symmetry restoration (CSR) phase transition (PT)  and the deconfinement one of color degrees of freedom are two different PTs or a single one.  On the other hand, a few advances  approaches  
support an idea that at finite values of  baryonic chemical potential a  phase with a partial CSR  occurs  before 
the deconfinement PT \cite{BugaevR1,BugaevR2,BugaevR3}.   However, one of  the hardest   problem of QCD phenomenology  is to determine the collision energy threshold of these PTs using the existing experimental data obtained in the central nucleus-nucleus collisions.

\begin{table}[t]
\caption{The summary of possible PT signals. The column II gives short description of the signal, while the columns III and IV 
indicate  its location,  status and references.}
\begin{center}
\footnotesize 
\begin{tabular}{l r r r }
\hline
~~~No and  Type  &   Signal \hspace*{28mm} &  C.-m.  energy $\sqrt{s}$ (GeV) &  C.-m.  energy $\sqrt{s}$ (GeV)  \\ 
& & Status &  Status \\ \hline

1.  Hydrodynamic     & Highly correlated  &    Seen at &  Seen at \\
 & quasi-plateaus in ent-&  {\bf 3.8-4.9 GeV} \cite{Bugaev:2014,Bugaev:2015}.       & {\bf 7.6-9.2 GeV}  \cite{Bugaev:2014,Bugaev:2015}.       \\ 
& ropy/baryon, ther- & Explained by the shock  &    \\
& mal pion number/ba-    &   adiabat  model  \cite{Bugaev:2014,Bugaev:2015}.  &  \\
 & ryon and total pion &   &  Require an explanation. \\
 &   number/baryon.  Sug- & & \\
&  gested in \cite{KAB:89a,KAB:91}.  &  &   \\ \hline

2. Thermodynamic& Minimum of the   &  In the one component &   \\ 
&chemical freeze-out  & HRGM it is seen &   \\
& volume $V_{CFO}$.  & at {\bf 4.3-4.9 GeV} \cite{Andronic:05}.  &  Not seen. \hspace*{11mm}  \\
& & In the multicomponent &   \\
& &  HRGM it is seen   &   \\
& &  at {\bf 4.9 GeV} \cite{Oliinychenko:12}.  &   \\
& &   Explained by the shock  &   \\
& & adiabat   model  \cite{Bugaev:2014,Bugaev:2015}.  &   \\ \hline

3. Hydrodynamic & Minimum of the  &  Seen at {\bf 4.9 GeV} \cite{Bugaev:2014}. &  Seen at {\bf 9.2 GeV} \cite{Bugaev:2014}.  \\
& generalized specific   &  Explained by the shock&   \\
&volume   $X = \frac{\epsilon +p}{\rho_b^2}$  at&  adiabat   model  \cite{Bugaev:2014,Bugaev:2015}.  & Require an explanation    \\ 
&  chemical freeze-out. & & \\ \hline

4. Thermodynamic&  Peak of the  trace & Strong peak is seen &  Small peak is seen \\ 
& anomaly $\delta = \frac{\epsilon-3p}{T^4}$.& at {\bf 4.9 GeV} \cite{Bugaev:2015}.  &  at {\bf 9.2 GeV} \cite{Bugaev:2015}.  \\
& &  Is generated &   \\
& &  by  the $\delta$ peak &   Require an explanation \\
& &   on the shock adiabat &   \\  
& &   at  high density end of &   \\  
& &    the mixed phase  \cite{Bugaev:2015}. &   \\  \hline

5. Thermodynamic&  Peak of the  bary- & Strong peak is seen &  Strong peak is seen  \\ 
& onic density $\rho_b$. & at {\bf 4.9 GeV} \cite{Bugaev17}.  &  at {\bf 9.2 GeV} \cite{Bugaev17}.  \\
& &  Is explained&    \\  
& &    by $\min\{V_{CFO}\}$ \cite{Oliinychenko:12}.  &  Require an explanation \\  \hline
  
6. Thermodynamic& Apparent chemical & $\gamma_s=1$ is seen &  $\gamma_s=1$ is seen at {\bf$\sqrt{s}$}  \\ 
  & equilibrium of  &  at {\bf 4.9 GeV} \cite{Bugaev17}.&  {\bf $ \ge$  8.8 GeV} \cite{Bugaev17,Andronic:05}.   \\
& strange charge.  & Explained by ther- &  Explained by ther- \\  
& & mostatic  properties &  mostatic  properties    \\  
& & of mixed phase  & of  QG bags with   \\
& & at $p=const$ \cite{Bugaev17}.    &  Hagedorn  mass   \\   
& & & spectrum   \cite{Bugaev17}. \\ \hline

7. Fluctuational &  Enhancement of &   &  Seen  at {\bf 8.8 GeV} \cite{KABKo2017}. \\
(statistical& fluctuations  &   N/A \hspace*{18mm} &  Can be explained by \\
mechanics)& &  &  CEP \cite{KABKo2017} or  3CEP\\  
& &  &  formation \cite{Bugaev17}.   \\   \hline

8. Microscopic  & Strangeness Horn &   &   Seen at {\bf 7.6 GeV}. Can \\
& ($K^+/\pi^+$ ratio)  &   N/A \hspace*{18mm}  &  be explained by the on- \\  
& &  &   set of  deconfinement  at\\
  & &  &  \cite{KABNayak2010}/above \cite{KABCassing16b}  {\bf 8.7 GeV}.  \\    \hline
\end{tabular}
\end{center}
\end{table}

Fortunately,  during last few years an essential progress in resolving such a problem was achieved \cite{Bugaev:2014,Bugaev:2015,Bugaev:2016a,KABCassing16a,KABCassing16b,KABKo2017}.  In particular,  two sets of remarkable hydrodynamic and thermodynamic signals of two PTs  at the center of mass collision energies $\sqrt{s_{NN}} = 4.3-4.9$ GeV and  $\sqrt{s_{NN}} = 7.6-9.2$ GeV were found in \cite{Bugaev:2014} and the hypothesis of their possible observation  at these energies of collision was first formulated in \cite{Bugaev:2014,Bugaev:2015,Bugaev:2016a}.  In the works \cite{KABCassing16a,KABCassing16b} a very good description of the large massive of experimental data on nuclear collisions  was first achieved with the Parton-Hadron-Sring-Dynamics (PHSD) model by assuming an existence of CSR PT at  about  $\sqrt{s_{NN}} \simeq 4$ GeV  in a hadronic phase and  a deconfinement one at $\sqrt{s_{NN}} \simeq 9-10$ GeV.  In 2017 the group of  scientists  analyzed the fluctuations of light nuclei and came to a conclusion that the vicinity of collision energy   $\sqrt{s_{NN}} \simeq 8.8$ GeV is a nearest vicinity of the critical endpoint of the QCD phase diagram \cite{KABKo2017}. The arguments of such a statistical signal of the  endpoint   were  essentially enhanced in  \cite{Bugaev17} and the conclusion of the tricritical endpoint existence  in QCD
at  or slightly above  $\sqrt{s_{NN}} \simeq 8.7-9.2$ GeV  was first formulated  in  \cite{Bugaev17}. { The summary of found signals is given in Table 1.}

Although the PHSD model provides us with some hints about the properties of the phase existing at the collision energy range 
 $\sqrt{s_{NN}} \simeq 4.9-9.2$ GeV, the question is whether one can  independently get the properties of this matter. In this work we briefly
 show how one can get them from the equation of state (EoS) which is obtained in  \cite{Bugaev:2014,Bugaev:2015} from fitting the data. 
 
 \section{Hadron Resonance Gas Model with Hard-Core Repulsion}
 
The possible PTs signals 1-6  presented in Table 1 were obtained with the help of the multicomponent Hadron Resonance Gas Model  (HRGM) \cite{Bugaev:2014,Bugaev:2015,Bugaev:2016a, Oliinychenko:12, KABSFO,KABVeta14, Bugaev:2016,KABSagun17}, which, in contrast to 
the HRGM with one or two hard core radii of hadrons \cite{Andronic:05} has the following hard-core radii  of pions
$R_{\pi}$=0.15 fm, kaons $R_{K}$=0.395 fm,  $\Lambda$-hyperons $R_{\Lambda}$=0.085 fm, other
baryons $R_{b}$=0.365 fm and  other mesons $R_{m}$=0.42 fm.  Thus, having only 2 or 3 additional   global fitting parameters compared 
to the usual  HRGM  \cite{Andronic:05},  one can  get extremely good description of the 
hadronic multiplicity ratios measured at AGS, SPS and RHIC energies with a  high quality $\chi^2/dof  \simeq 1.04$  \cite{Bugaev:2016a,KABSagun17}, including traditionally  the 
most problematic ones for the usual HRGM  \cite{Andronic:05}, i.e.  $K^+/\pi^+$, $\Lambda/\pi^+$ and $\bar \Lambda/\pi^-$ ratios.

A high quality fit of hadronic multiplicity ratios achieved by  the multicomponent HRGM gives us a high confidence 
that the EoS of hadronic matter is now fixed with high accuracy
in the wide range of  chemical freeze-out (CFO) temperature $T$ and baryonic chemical potential $\mu_B$.  This conclusion was thoroughly verified 
recently  with  the newest version \cite{Bugaev:2016,KABSagun17}  of the multicomponent HRGM which allows one to  go beyond the Van der Waals approximation traditionally used
in HRGM.  In the simplest case of a single hard-core radius of hadrons $R$ the HRGM pressure in the grand canonical ensemble  is
\begin{eqnarray}
\label{EqI}
p&=&\sum_n \, p_n^{id} \, (T, \mu_n-b\,p) \,, \quad p_n^{id} (T, \mu) = g_n  \int\frac{{d \bf k}} {(2\pi^3)} \frac{k^2} {3\, E_n (k)} \frac{1} { \exp{\left( \frac{E_n (k)  - \mu} {T} \right)} + \zeta_n}
\end{eqnarray}
where the sum  is  running  over all particles (and antiparticles)  with the chemical 
potentials  $\mu_n$,  $b = 4 V_0= 4\, \frac{4}{3}\pi R^3$ is the excluded volume of  hadrons and $V_0$ is their proper volume.
Here $p_n^{id}(T, \mu)$  denotes the partial pressure 
of the point-like hadrons  of sort $n$ with  the degeneracy $g_n$ and the mass $m_n$, while 
$E_n(k) = \sqrt{{\vec k}^2 + m_n^2}$ is the energy of particle with the 3-momentum $\vec k$  and $\mu$ is the effective chemical potential.  The parameter  $\zeta_n$ defines  the Fermi ($\zeta_n=1$), the Bose  ($\zeta_n=-1$) or  the Boltzmann  ($\zeta_n=0$)
 statistics. Then the thermal density of particles of sort $l$ is defined as
\begin{eqnarray} 
\label{EqII}
&&n_l \equiv \frac{\partial p}{\partial \mu_l} = \frac{n_l^{id} } {1 + b\, \sum_k n_k^{id}} \,, \quad 
n_l^{id} (T, \nu, \zeta_l) =  g_l  \int\frac{d{\bf k}} {(2\pi^3)}  \frac{1} { \exp{\left( \frac{E_l (k)  - \nu} {T} \right)} + \zeta_l}
 \,,
\end{eqnarray}
where  $n_l^{id} (T, \nu, \zeta_l)$ denotes the particle number density of  point-like hadrons of  sort $l$.

 \section{Necessity  of  Multicomponent HRGM}
There are two main reasons of why the HRGM of Eqs. (\ref{EqI})  and (\ref{EqII})  with the Van der Waals  is used to determine the CFO parameters. The main reason is that for the hard-core repulsion the energy per particle is same as in the ideal gas and, therefore, there is no need
to transform the potential energy of the system into the kinetic energy of particles. Of course, one could add the attractive term $P_{attr} (\{n_k\})$ to the pressure 
(\ref{EqI}), but in this case one would face a hard mathematical  problem to convert the interacting gas into free streaming hadrons \cite{KABkinFO1,KABkinFO2,KABkinFO3} measured in the experiments. 

The second reason is that only the hard-core repulsion provides the  consistence with the lattice QCD results. In other words, if one takes into account all hadrons as  the 
point-like particles with $b=0$, then it is well-known that at high $T$ and $\mu_B$ their  pressure  will dramatically exceed the  pressure of quarks and gluons. However, in order to provide a high quality fit of the data such a simplified model should be modified in two respects.  First of all
one should remember that the quantum second  virial coefficients of  particle of sort $k$ interacting with the particle of sort $l$ is \cite{BugaevQGAS}
\begin{eqnarray}
\label{EqIII}
&&\hspace*{-0.9cm}a_{2, kl}^{Q} = b + a_{2, k}^{(0)} \delta_{kl} - \frac{a^{attr}_{kl}}{T} = b + \frac{\zeta_k}{2}  \frac{n_k^{id} (T/2, 0, 0) } {\left[ n_k^{id}(T, 0, 0)  \right]^{2}}  \delta_{kl}  - \frac{1}{2\, T} \lim_{\{n_m\} \rightarrow 0} \frac{\partial^2 P_{attr} (\{n_m\})}{\partial n_k\,  \partial n_l} \,,
\end{eqnarray}
where the term $a_{2, k}^{(0)}$ is the  virial coefficient due to quantum statistics of hadron of sort $k$  which is expressed in terms of 
the densities $n_k^{id}(T, 0, 0)$ of  auxiliary  Boltzmann hadrons  of  the same sort $k$, and the term $a^{attr}_{kl}$ is due to attractive interaction.  This equation shows that the  gas pressure (\ref{EqI}) with the hard-core repulsion, indeed, accounts for the quantum properties of hadrons, if 
$\zeta_k\neq 0$. It also  shows
 that, if one introduces the different hard-core radii of hadrons, then one  can even account for the attraction between them at the level of the second virial coefficient which is sufficient for the low particle densities at CFO. Of course,  for all hadrons the second virial coefficients  (\ref{EqIII}) are temperature dependent, but fortunately, at high temperatures such a dependence is not strong \cite{BugaevQGAS} and, hence, to a leading order one can restrict the treatment by the constant hard-core radii. 

In addition, the HRGM pressure corresponds to the hadron resonances of vanishing width. This is, of course, a rough approximation because 
 at  CFO the density is sufficiently low that the inelastic reactions between  hadrons can be neglected and, hence, the hadrons and their resonances  should get  their vacuum masses and vacuum widths before going into detector.  The other reason to introduce the widths is the practical one. Thus, using the 
Briet-Wigner parameterization of resonance width of all hadronic resonances    one can describe the hadron rations essentially better
than with the Gaussian one or with the vanishing width  \cite{BugaevWidth}. Therefore,  it seems that  the most efficient way to 
account for the residual attractive interaction between hadrons at CFO and to achieve a high quality of the hadron multiplicity description
is to generalize  the one component HRGM  (\ref{EqI}),  (\ref{EqII})  to the multicomponent case, i.e. to account for different hard-core radii
and then  to determine these radii  from the fit of experimental data. This is exactly what was done in Refs. \cite{Bugaev:2014,Bugaev:2015,Bugaev:2016a, Oliinychenko:12, KABSFO,KABVeta14, Bugaev:2016,KABSagun17,BugaevWidth} during last five years.
It is evident that the hard-core radii determined in this way are effective ones by  construction. 
  
 \section{EoS of Hadronic  Matter with CSR}
 
 Using the multicomponent HRGM in Refs. \cite{Bugaev:2014,Bugaev:2015} it was possible 
   from fitting the entropy per baryon  $s/\rho_B$ along the  shock adiabat \cite{KAB:89a,KAB:91} to determine the  EoS of the phase existing at 
 the collision energy range $\sqrt{s_{NN}} \simeq 4.9-9.2$ GeV.  This   EoS is   similar  to the MIT-Bag model
\begin{eqnarray}\label{EqIV}
p_{Chiral} =A_0T^4+A_2T^2\mu^2+A_4\mu^4-B  \,, 
\end{eqnarray}
but the coefficients  $A_0 \simeq 2.53 \cdot 10^{-5}~{\rm MeV}^{-3}{\rm fm}^{-3}$, $A_2 
\simeq 1.51 \cdot 10^{-6} ~{\rm MeV}^{-3}{\rm fm}^{-3}$, $A_4 \simeq 1.001 \cdot 10^{-9}~{\rm 
MeV}^{-3}{\rm fm}^{-3}$, and $B \simeq 9488~{\rm MeV}~{\rm fm}^{-3}$ are rather different 
from what is predicted by the perturbative QCD for massless gluons and  (anti)quarks.  In Ref. \cite{Bugaev17} the EoS (\ref{EqIV}) was suggested  to   find  out  the number of  bosonic and fermionic degrees of freedom of this phase.  
Recalling that  first three terms of the EoS (\ref{EqIV}) correspond to the gas of  massless particles and noting that the coefficient  $A_4 $ is small and its value is comparable to its own error, we could determine the numbers of total $N_{dof}^{tot}$,  bosonic $N_{b}^{eff}$  and fermionic $N_{f}^{eff}$ degrees of freedom as 
\begin{eqnarray}\label{EqV}
N_{dof}^{tot} =  \frac{90}{\pi^2} \, A_0 \hbar^3  \simeq 1770   \,, \quad  N_{f}^{eff} =  12 \, A_2 \hbar^3  \simeq  141 \,, \quad  N_{b}^{eff} =
N_{dof}^{tot} -  \frac{7}{4} N_{f}^{eff} \simeq  1523 \,. 
\end{eqnarray}
Since the numbers  $N_{b}^{eff}$ and $N_{f}^{eff}$ are much larger than the corresponding number of degrees of freedom in perturbative QCD, but at the same time  they are close to the  total number of spin-isospin degeneracies of all known hadrons, in  Ref. \cite{Bugaev17} we, independently 
of the works \cite{KABCassing16a,KABCassing16b}, concluded that the EoS (\ref{EqIV}) corresponds to the gas of massless hadrons with strong attraction given by the vacuum pressure $B$. 
\section{Conclusions}
Here we present a  summary of   possible signals of CSR and deconfinement PTs  which may be, respectively,  probed at 
the collision energies at  $\sqrt{s_{NN}}\simeq  4.3-4.9$ GeV  and  above $\sqrt{s_{NN}}\ge  8.7-9.2$ GeV. 
Also these signals may 
evidence for an existence of the 
tricritical endpoint of QCD phase diagram  at the collision energy around  or slightly above  $\sqrt{s_{NN}}\ge  8.7-9.2$ GeV. 
The EoS of the hadronic matter with CSR is discussed and the number of bosonic and fermionic degrees of freedom is found.

\ack 
K.A.B., A.I.I., V.V.S. and G.M.Z.  acknowledge  a partial  support from 
the program ``Nuclear matter under extreme conditions'' launched 
by the Section of Nuclear Physics of  the National Academy of Sciences  of Ukraine. 
 The work of K.A.B. and L.V.B. was performed in the framework of COST Action CA15213 ``Theory of hot matter and relativistic heavy-ion collisions" (THOR). K.A.B. is thankful to the  COST Action CA15213  for a partial support. 
 V.V.S. acknowledges a partial support by grants from ``Funda\c c\~ao para a Ci\^encia e Tecnologia".
 

\end{document}